\documentclass[12pt,preprint]{aastex}

\shorttitle{COSMOLOGICAL CONSTRAINTS ON HIGGS BOSON MASS}
\shortauthors{L.A. Popa et al. }
\begin{document}

\title{COSMOLOGICAL CONSTRAINTS ON THE HIGGS BOSON MASS}
\author{L.A. Popa, A. Caramete}
\affil{ Institutul de \c{S}tiin\c{t}e Spa\c{t}iale Bucure\c{s}ti-M\u{a}gurele, Ro-077125 Rom\^{a}nia}

\email{lpopa@venus.nipne.ro }

\begin{abstract}

For a robust interpretation of upcoming observations
from PLANCK and  LHC  experiments
it is imperative to understand how the inflationary dynamics of a
non-minimally coupled Higgs scalar field with gravity may affect
the determination of the inflationary observables.
We make a full proper analysis of the WMAP7+SN+BAO dataset
in the context of the non-minimally coupled  Higgs inflation
field with gravity.

For the central value of the top quark pole mass $m_T=171.3$GeV,
the fit of the inflation model with non-minimally coupled Higgs scalar field
leads to the Higgs boson mass in range
$143.7 \,{\rm GeV} \le m_H \le 167 {\rm GeV}$ (95\% CL)\\
We show that the inflation driven by a non-minimally coupled scalar field to
the Einstein gravity leads to significant constraints on the scalar spectral index $n_S$ and
tensor-to-scalar ratio $R$ when compared with  a tensor with similar constraints to
form the standard inflation with a minimally coupled scalar field. \\
We also show that an accurate reconstruction of the Higgs potential in terms of
inflationary observables requires an improved accuracy of other
parameters of the Standard Model of particle physics such as the top quark mass and the effective QCD coupling constant.

\end{abstract}


\section{Introduction}
\label{S1}

The primary goal of particle cosmology is to obtain a concordant description of
the early evolution of the universe, establishing a testable link between cosmology
and particle physics, consistent with both unified field theory
and astrophysical and cosmological measurements.
On the ground, the Large Hadron Collider (LHC) at CERN is investigating the elementary
particle collisions in the TeV energy range,
seeking to validate a large number of theoretical predictions of the Standard Model (SM) of particle physics and beyond. In the sky, the PLANCK
Surveyor is actively taking precise measurements of the Cosmic Microwave Background (CMB) temperature and
polarization anisotropies.

Inflation is the most simple and robust theory capable of explaining astrophysical
and cosmological observations, at the same time providing self-consistent primordial initial conditions \cite{Staro80,Guth81,Sato81,Albercht82,Linde82,Linde83} and mechanisms for
the quantum generation of scalar (curvature) and tensor (gravitational waves)
perturbations \cite{Muk81,Haw82,Guth82,Staro82,Bardeen83,Abbot84}.
In the simplest class of inflationary models, inflation is driven by a single
scalar field $\phi$ (or inflaton) with some potential $V(\phi)$ minimally coupled to the Einstein gravity. The perturbations are predicted
to be adiabatic, nearly scale-invariant and Gaussian distributed,
resulting in an effectively flat universe.\\
The WMAP cosmic microwave background (CMB) measurements alone \cite{Dunkeley,Larson} or
complemented with other cosmological datasets \cite{Komatsu,Komatsu10}
support  the standard inflationary predictions of a nearly flat universe with adiabatic initial density perturbations. In particular, the detected anti-correlations between temperature and E-mode polarization
anisotropy on degree scales \cite{Nolta} provide strong evidence for correlation on
length scales beyond the Hubble radius.

Alternatively, one can look to the inflationary dynamics based on models beyond the Standard Model (SM) of particle physics. The hybrid inflation models involving supersymmetric (SUSY) TeV energy scales   \cite{Dvali94}  and minimal supergravity (SUGRA) \cite{Linde94} provide natural connection between cosmology and particle physics \cite{Cervantes95, Senoguz03}.
The realization of these inflationary scenarios introduces new physics between the electroweak energy scale and the Planck scale, leading to distinct predictions of the main inflationary parameters, such as the spectral index $n_S$ of scalar perturbations
and the tensor-to-scalar ratio $R$ \cite{Rehman08,Rehman09,Rehman10}.\\
However, a number of recent papers \cite{BezSha08,BarSta08,BerSha09,BerSha09,deSimone09,BGS09}
reported the  possibility that the SM of particle physics with an additional non-minimally coupled term of the Higgs field
to the gravitational Ricci scalar  can give rise to inflation without the need for additional degrees of freedom to the SM.
This scenario is based on the observation that the problem of the
very small value of Higgs quadratic coupling required  by the CMB anisotropy data
can be solved if the Higgs inflaton has a large coupling with gravity
\cite{Futamase89,Fakir90,Komatsu99,Tsu04,BarKa94}. \\
The resultant Higgs inflaton  effective  potential
in the inflationary domain is effectively flat and can result in a successful inflation for values of the non-minimal coupling constant  $\xi \sim 10^3 - 10^4$, allowing for cosmological values for the Higgs boson mass in a window in which the electroweak vacuum is stable and therefore sensitive to  the field fluctuations during the early stages of the universe \cite{Espinosa08}.

Limits of the validity of Higgs-type inflation have recently been debated by several
authors.
Specifically, Barb\'{o}n \& Espinosa (2009) argued that
the large coupling of Higgs inflaton to the Ricci scalar
makes this model invalid beyond the ultraviolet cutoff scale
$\Lambda_{\xi} \simeq M_{P}/\xi$ (here $M_P=2.4 \times 10^{18}$ GeV is the reduced Planck mass) which is below the Higgs field expectation value  at $N$ {\it e}-foldings during inflation, $h \simeq \sqrt{N}M_P/\sqrt{\xi}$.
As consequence,
at the ultraviolet cutoff scale  $\Lambda_{\xi}$
at least one of the cross-sections of different scattering processes hits the unitarity bound \cite{Burgess09}.
The fact that the quantum corrections due to the strong
coupling to gravity makes the perturbative analysis to break down at energy scales above $\Lambda_{\xi}$ was interpreted as a signature of a new physics, implying higher dimensional operators at energies above $\Lambda_{\xi}$.
However, the theory can still be considered valid above
$\Lambda_{\xi}$ if one finds some ultraviolet completion
or if a very high degree of fine tuning is required, keeping in this way the unwanted contributions of higher dimensional operators
small to zero \cite{BezSha09,deSimone09}. \\
Recent papers
\cite{Lerner10a,Lerner10b,Burgess10,Hertzberg10}
revisit the arguments against Higgs-type inflation addressing the issue of its
naturalness with respect to perturbativity and unitarity violation in
the Jordan and Einstein frames.
It is shown that
the apparent breakdown of this theory in the Jordan frame does not imply new physics, but a failure of the perturbation theory in the Jordan frame
as a calculational method.
These works demonstrate that for inflation based on a single scalar field
with large non-minimal coupling, the quantum corrections at high energy scales
are small, making the perturbative analysis valid.  As consequence, for these models there is no breakdown of unitarity at the energy scale $\Lambda_{\xi}$.
In particular, when the single-field Higgs inflation model is analyzed in the Einstein frame there is no breakdown of the theory at energy scales ${\hat h} \geq \Lambda_{\xi}$, where ${\hat h} $ is the canonically normalized Higgs scalar field in the Einstein frame.
However, the inclusion of two or more scalar fields non-minimally coupled with gravity (in particular, the 3 Goldstone bosons of the Higgs doublet)
causes unitarity violation  in the Einstein frame  at  $\Lambda_{\xi}$,
making the theory unnatural \cite{Hertzberg10}.

The present cosmological constraints on the Higgs mass
are based on mapping between the Renormalization Group (RG) flow equations and
the spectral index of the curvature perturbations parameterized in terms of the number of ${\it e}$-foldings until the end of inflation,  emerging from the analysis of CMB data combined with astrophysical distance measurements.
For a robust interpretation of upcoming observations
from PLANCK \cite{Planck}  and  LHC \cite{LHC} experiments it is imperative to understand how the inflationary dynamics of a non-minimally coupled Higgs scalar field may affect  the degeneracy of the inflationary observables. \\
The aim of this paper is to make a full proper analysis of the WMAP 7-year
CMB measurements complemented  with astrophysical distance measurements
\cite{Komatsu10,Larson} in the context of the non-minimally coupled  Higgs inflaton field with gravity.
The paper is structured as follows.
In Section~2 we compute the power spectra of
scalar and tensor density perturbations generated during inflation
driven by a single scalar field non-minimally coupled to gravity.
In Section~3 we derive the Higgs field equations
and compute the RG improved Higgs field potential and in Section~4 we present
our main results. In Section~5 we draw our conclusions.
Throughout the paper $a$ is the cosmological scale factor ($a_0=1$ today),
$\kappa^2 \equiv 8 \pi M_{pl}^{-2}$ where $M_{pl}\simeq 1.22 \times 10^{19}$ GeV is the present value of the Planck mass, overdots denotes the time derivatives and
$_{,\varphi} \equiv \partial/\partial \varphi $. \\

\section{COSMOLOGICAL PERTURBATIONS DRIVEN BY A NON-MINIMALLY COUPLED SCALAR FIELD}
\label{S2}

In this section we compute the power spectra of
scalar and tensor density perturbation generated during inflation
driven by a single scalar field non-minimally coupled to gravity via
the Ricci scalar  \cite{Fakir90,Noh96,Komatsu98,Komatsu99,Noh01,Tsu04} .
The general action for these models in the Jordan frame is given by \cite{Futamase89}:
\begin{equation}
\label{action}
S_{J} \equiv \int d^4 x \sqrt{-g} \left[  U(\varphi) {\it \bf R}
            -\frac{1}{2}G(\varphi)(\nabla \varphi)^2 -V(\varphi)\right] \,,
\end{equation}
where $U(\varphi)$ is a general coefficient of the Ricci scalar, {\it \bf R}, giving rise to the non-minimal coupling, $G(\varphi)$ is the general coefficient of kinetic energy and $V(\varphi)$
is the general potential. \\
The generalized $U(\varphi)${\it \bf R} gravity theory in Equation (\ref{action})
includes diverse cases of coupling. For the generally coupled scalar field
$U=(\gamma + \kappa^2\xi \varphi^2)$,  $G(\varphi)=1$ and
$\gamma$ and $\xi$ are constants. The non-minimally coupled scalar field is the case
with $\gamma=1$ while the conformal coupled scalar field is the case with
$\gamma=1$ and $\xi=1/6$. \\
The conformal transformation for the action given  in Equation (\ref{action}) can be achieved by defining the Einstein frame metric as:
\begin{eqnarray}
\label{CT}
\hat {g}_{\mu,\nu}= \Omega {g}_{\mu,\nu}\,,\hspace{0.3cm}\hspace{0.3cm}
\Omega=2 \kappa^2 U(\varphi) \,,
\end{eqnarray}
where the quantities in the Einstein frame are marked by caret.

The kinetic energy in the Einstein frame can be made canonical with respect to
the new scalar field ${\hat \varphi}$, defined through the scalar field propagator suppression factor $s(\hat{\varphi})$ as \cite{deSimone09,BarSta09}:
\begin{equation}
\label{dphi}
s(\hat{\varphi})^{-2}=\left( \frac{{\rm d}\hat{\varphi}}{{\rm d}\varphi}\right)^2  =
\frac{1}{2 \kappa^2}\frac{G(\varphi) U(\varphi)+3 U^2(\varphi)_{,\varphi}}
{  U^2(\varphi)} \,.
\end{equation}
Thus the non-minimal coupling to the gravitational field introduces a modification to the Higgs field propagator by the factor $s(\hat{\varphi})$, acting as {\it back reaction} of
the gravitational field. \\
The scalar potential $\hat{V}(\hat{\varphi})$ in the Einstein frame
is given by:
\begin{equation}
\label{Ve}
\hat{V}(\hat{\varphi}) =  \frac{1}{4 \kappa^4}\frac{V(\varphi)}{U^2(\varphi)}\,,
\end{equation}
leading to the following canonical form of the action in the Einstein frame:
\begin{equation}
\label{action_new}
S_{E} \equiv \int d^4 x \sqrt{-g} \left[ \frac{1}{2\kappa^2}  {\it \bf R}
 -\frac{1}{2}(\nabla \hat{\varphi})^2 -V(\hat{\varphi})\right] \,.
\end{equation}

\subsection{Background Field Equations}

When evaluating the field equations we assume that the background
space-time can be written in the
form of a flat (k=0) Robertson-Walker line element:
\begin{eqnarray}
\label{RW}
{\rm d}s^2=g_{\mu,\nu}\,{\rm d}x_{\mu}\,{\rm d}x^{\nu} &  = & -{\rm d}t^2+a^2(t){\rm d}x^2 \\ \nonumber
& = & \Omega(x) \left(-{\rm d}{\hat t}^2+\hat{a}^2({\hat t}){\rm d}x^2\right)\,,
\end{eqnarray}
where $t$ is the cosmic time and $a$ is the cosmological scale factor.
From the above equation we obtain:
\begin{eqnarray}
\label{ta}
{\rm d}\hat{a}=\sqrt{\Omega}\,{\rm d}a\,, \hspace{1cm}
{\rm d}\hat{t}= \sqrt{\Omega}\, {\rm d}t \,.
\end{eqnarray}
Now the Friedmann equation in the Einstein frame can be written as \cite{Komatsu98,Tsu04}:
\begin{equation}
\label{ein_eq}
\hat{H}^2  =  \frac{\kappa^2}{3}\left[\left(\frac{{\rm d}\hat{\varphi}}{{\rm d}\hat{t}}\right)^2 +
\hat{V}(\hat{\varphi})\right]\,, \\
\end{equation}
where:
\begin{eqnarray}
\label{hub}
\hat{H}  \equiv  \frac{1}{\hat{a}}
\frac{ {\rm d}\hat{a}} {{\rm d}\hat{t}} &  = &  \frac{1}{\sqrt{\Omega}}
\left[ H+ \frac{1} {2 \Omega} \frac{ {\rm d} \Omega} { {\rm d} t } \right]\,,\\
\label{field}
\frac{ {\rm d}\hat {\varphi}} {{\rm d} \hat{t}}  & = &
\left( \frac{ {\rm d} \hat{\varphi}} { {\rm d} \varphi } \right)
\left( \frac{ {\rm d} t } { {\rm d} \hat{t} } \right) \dot{\varphi}
\end{eqnarray}
Equations (\ref{hub}) and (\ref{field}) are enough to compute the background
field evolution in the Einstein frame if the field equations in the Jordan
frame are known (see the next section).

\subsection{Scalar and Tensor Perturbations}

Neglecting the contribution of the decaying modes,
the scale dependence of the amplitudes of scalar (S) and tensor (T)
perturbations in the Einstein frame are fully governed by
the mode  equation \cite{Muk81}:
\begin{eqnarray}
\label{Muk}
\frac{{\rm d}^2 u_k} {{\rm d}{\hat t}^2}+ \left(k^2- \frac{1}{{z}} \frac{{\rm d}^2 z}{{\rm d}{\hat t}^2} \right)u_k  =  0 \,,
\end{eqnarray}
where $k$ is the comoving wave number of the  mode function $u_k$.
For the case of scalar perturbations we have \cite{Hwang96,Noh96,Noh01}:
\begin{eqnarray}
\label{zS}
\frac{1}{z_S} \frac{{\rm d}^2 z_S}{{\rm d}{\hat t}^2}=
({\hat a}{\hat H})^2 \left[ (1+{\hat \delta}_S) (2+{ \hat \delta}_S + \hat{\epsilon} ) +
\frac {\dot {\hat \delta}_S}{ {\hat a}{\hat H}  } \right]\,,
\end{eqnarray}
where:
\begin{eqnarray}
z_S=\hat{a} \sqrt{ {\hat{Q}_S } } \,, \hspace{1cm}
\hat{Q_s}= \left( \frac{ {\rm d} \hat{\varphi} / {\rm d} \hat{t} } {\hat{H}} \right)^2 \,.
\end{eqnarray}
and slow-roll parameters $\hat{\epsilon}$ and $\hat{ \delta}_S$ are given by \cite{Stew93}:
\begin{eqnarray}
\hat{\epsilon}=   -\frac{ {\dot {\hat H}} }{\hat{H}^2}\,,
\hspace{1cm}
\hat{ \delta}_S=  \frac{ {\dot {\hat Q}}_S} {2 \hat{H} \hat{Q}_S } \,.
\end{eqnarray}
In the case of tensor perturbations Equation~(\ref{zS})
has the same  form with the following replacements:
\begin{eqnarray}
z_S \rightarrow z_T=\hat{a} \sqrt{ {\hat{Q}_T } } \,, \hspace{1cm}
{\hat Q}_S \rightarrow {\hat Q}_T =1 \,,
\hspace{1cm}
\hat{ \delta}_S \rightarrow  {\hat \delta_T}=
\frac{ {\dot {\hat Q}}_T} {2 \hat{H} \hat{Q}_T }=0 \,.
\end{eqnarray}
The power spectra of scalar and tensor perturbations are given by \cite{Copeland94}:
\begin{eqnarray}
\label{As_e}
{\cal P}_S (k)= \frac{k^3}{2 \pi^2} \, \left ( \frac{1}{Q_S}\right)^2 \,\frac{|u_k|^2}{a^2}
\hspace{0.8cm}
\label{At_e}
{\cal P}_T (k)= \frac{16 k^3}{\pi m^2_{pl}} \,  \frac{|u_k|^2}{a^2}  \,,
\end{eqnarray}
and the spectral index of the scalar perturbations $n_{S}$
is obtained as usual as: $n_s-1=  d\,{\rm ln} {\cal P}_S(k)/ d\,{\rm ln} k$.

\section{HIGGS BOSON AS INFLATON}
\begin{figure}
\label{rg_couplings}
\epsscale{1.}
\plotone{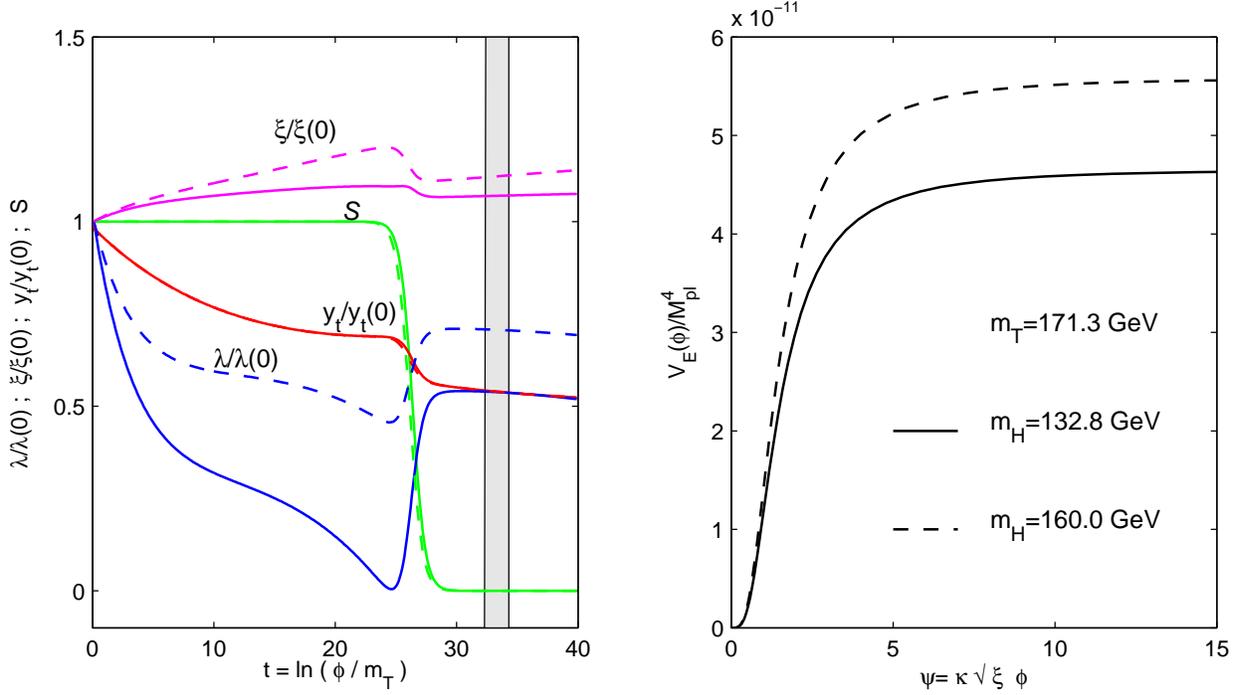}
\caption{Left panel: The running of the coupling constants
normalized to their initial values for
$m_{H}=132.8$ GeV (continuous lines) with $\lambda(0) \simeq 0.14$ (blue), $\xi(0)=1.7 \times 10^{4}$ (magenta), $m_{H}=160$ (dashed lines) with $\lambda(0)=0.21$ (blue),  $\xi(0)=2.1 \times 10^4 $ (magenta) and  $m_{T}=171.3$ GeV with $y_t(0)= 0.91 $ (continuous red line).
The green curves show the running of the Higgs field propagator suppression factor $s(t)$.
The right-hand gray region indicates the slow-roll inflationary regime.
Right~panel:~The~Einstein~frame~renormalization~group~improved~potential~as~a~function~of~the  Higgs field for $m_{H}=132.8$ GeV (continuous line), $m_{H}=160$ (dashed line)  and $m_{Top}=171.3$ GeV.
In both cases we take the amplitude of scalar perturbations $A^2_S=2.44 \times 10^{-9}$ at the Hubble radius crossing $k_*$=0.002 Mpc$^{-1}$ and the vacuum expectation value ${\it v}$=246.22 GeV.}
\end{figure}
Higgs boson as inflaton adds non-minimal coupling to gravity
\cite{BarKa94,BezSha08,BarSta08,deSimone09,BerSha09,BezSha09}. \\
Taking the Higgs field potential $V(\varphi)$ of the Landau-Ginzburg type
(by assuming that the spontaneous symmetry breaking arises through a condensate),
the Jordan-frame effective action
has the same form as given in Equation (\ref{action})
with (see e.g. Futamase \& Maeda 1989, Fakir \& Unruh 1990, Makino \& Sasaki 1991):
\begin{eqnarray}
\label{UGV}
U(\varphi)= \frac{1+\kappa^2 \xi \varphi^2}{2 \kappa^2} \,,
\hspace{1.5cm}
V(\varphi) =\frac{\lambda}{4} \left(\varphi^2-{\it v}^2 \right)^2 \,,
\hspace{1.5cm}
G(\varphi)=1\,,
\end{eqnarray}
where: ${\it v}=(\sqrt{2}G_F)^{-1/2}$=246.22 GeV is the vacuum expectation value  of the Higgs field that sets the electroweak scale, $\lambda$ is
the quadratic coupling constat of the Higgs boson with a mass $m_H=\sqrt{2 \lambda}{\it v}$
and $\xi$ is the non-minimal coupling constant.
The Jordan-frame field equations from the above action are given by \cite{Komatsu99,Kaiser95}:
\begin{equation}
 H^2  =  \frac{\kappa^2}{3\,(1+\kappa^2\xi\phi^2)}
\left[V(\varphi) + \frac{1}{2} \dot{\varphi}^2- 6\,\xi H \varphi \dot{\varphi} \right]\,,
\end{equation}
\begin{eqnarray}
\ddot{\varphi}  + 3 H \dot{\varphi}  +  \left( \frac{ \kappa^2 \xi \varphi^2 (1+6 \xi) } {1+\kappa^2 \xi \varphi^2
 (1+6 \xi)}\right)  \frac{ \dot{\varphi}^2} {\varphi}  =
      \frac{ \kappa^2 \xi \varphi V(\varphi) - ( 1+ \kappa^2
      \xi \phi^2)V_{,\,\varphi}(\varphi)}{1+\kappa^2 \xi \varphi^2 (1+6\xi)} \,,
\end{eqnarray}
which in the slow-roll approximation
( $|\dot\varphi/\varphi| \ll H$ and $|\dot{\varphi}^2| \ll V(\varphi)$ ) can be written as:
\begin{eqnarray}
\label{E1}
H^2 & \simeq & \frac{\kappa^2}{3(1+\kappa^2\xi\phi^2)}V(\varphi) \,, \\
\label{E2}
3 H \dot{\varphi} & \simeq &
\frac{ \kappa^2 \xi \varphi V(\varphi) - ( 1+ \kappa^2 \xi \phi^2)V_{,\,\varphi}(\varphi) }
{1+ \kappa^2 \xi \varphi^2 (1+6\xi)} \,.
\end{eqnarray}
The quantum corrections due to the interaction effects of
the SM particles with Higgs boson through quantum loops modify
the action coefficients $U(\varphi)$, $V(\varphi)$ and $G(\varphi)$ from their classical expression
given in Equations (1) and (\ref{UGV}), taking the renormalization group (RG) improved
forms $U_q(t)$, $V_q(t)$, $G_q(t)$ defined as \cite{BarSta08,deSimone09,Clark09,Lerner09}:
\begin{eqnarray}
\label{UGV_rg}
U_q(t) & = & \frac{1}{2\kappa^2}
\left( 1 + \kappa^2 \xi(t)G_q^2(t)\varphi(t)^2\right) \,,\\
V_q(t) & = &  \frac{\lambda(t)} {4} G_q^4 (t)
\left(  \varphi^2(t)- {\it v}^2 (t) \right)^2 \,, \\
G_q(t) &  = & e ^{-\gamma(t)/(1+\gamma(t))} \,,
\end{eqnarray}
where $\gamma(t)$ is the Higgs field anomalous dimension given in the Appendix.  \\
The scaling variable $t= {\rm ln}(\varphi/m_T)$
in the above equations normalizes the Higgs field and all the running couplings to the top quark mass scale $m_T$.

As the energy scale of inflation is many order of magnitude above the electroweak scale ($\varphi(t) >> {\it v}$), in the following we will approximate the Higgs potential by $V(\varphi) \simeq \lambda \varphi^4/4$, neglecting the vacuum contribution and its running in the potential.\\
Making the conformal transformation (\ref{CT}),
Equations (\ref{dphi}) and (\ref{Ve}) yield to:
\begin{eqnarray}
\label{s_factor}
s(t)^{-2}=\left(\frac{{\rm d}\hat{\varphi}(t)} {{\rm d}\varphi(t)} \right)^2
& = & \frac{1}{2 \kappa^2} \,\frac{ 1+\kappa^2 \xi(t) \varphi^2(t)(1+6 \xi(t))}{(1+\kappa^2\xi(t) \varphi^2(t))^2}\,, \\
\label{V_E}
\hat{V}(t) & = & \frac{1}{16 \kappa^4} \,
\frac{\lambda(t)\varphi^4(t)}
{ (1+\kappa^2\xi(t)\varphi^2(t))^2}\,.
\end{eqnarray}
The amplitude of scalar density perturbations at the Hubble radius crossing $k_*$ is then given by:
\begin{eqnarray}
\label{A_S}
A^2_S=\left . \frac{ {\hat V}}{24 \pi^2 M^2_{pl} \epsilon} \right|_{k_*}\,,
\hspace{1.5cm}
\epsilon=\frac{1}{2}M^2_{pl}\left( \frac{\hat{V}_{\,,\varphi}}{\hat{V}} \right)^2 \,.
\end{eqnarray}
We compute the various $t$-dependent running constants,
the Higgs field propagator suppression factor and the Higgs field anomalous dimension  by integrating the RG  $\beta$-functions as compiled in the Appendix.
The runnings of $SU(2) \times S(1)$ gauge
couplings ${g',g}$, the $SU(3)$ strong coupling $g_s$,
the top Yukawa coupling $y_t$ and the Higgs quadratic coupling $\lambda$
are computed by using two-loop quantum corrections while
the running of non-minimal coupling constant $\xi$
is computed by using one-loop quantum corrections. \\
One should note the importance of the quantum corrections due to non-minimal coupling.
The quantum corrections to the classical kinetic sector $G(\varphi)=1$
 arise from the Higgs field
anomalous dimension $\gamma(t)$ occurring with a factor of $1/\xi$ which in the inflationary regime ($\xi \sim 10^4$) has a negligible small contribution.
In the case of a classical gravity sector
$U(\varphi)= (1+\kappa^2 \xi \varphi^2)/2 \kappa^2$, the conformal transformation (2)
introduces a one-loop $\beta$-function for $\xi$  with a term proportional to $\lambda$
due to Higgs running in a loop which has a small contribution during inflation due to the suppression
of the Higgs field propagator, while the contribution of the remaining terms cancel to good approximation \cite{deSimone09}.
Although small, the one-loop quantum corrections due to the non-minimal coupling
are not negligible but enough for the purpose of  this analysis.

For each case, the $t$-dependent running constants are obtained as:
\begin{eqnarray}
\label{running}
Y(t)=\int^{t}_{t=0} \frac{\beta_{Y}(t')}{1+\gamma(t')}\,{\rm d}t' \,,
\hspace{1.5cm} Y=\{g,g',g_s,y_t,\lambda,\xi \}\,,
\end{eqnarray}
At $t=0$, which corresponds to the top quark mass scale $m_T$,
the Higgs quadratic coupling $\lambda(0)$ and  the top Yukawa coupling $y_t(0)$  are determined by the pole masses and the vacuum expectation value ${\it v}$:
\begin{eqnarray}
\label{pol_mass}
\lambda(0)=\frac{m^2_H}{2 {\it v}^2}\left[1+ \Delta_H (m_H)\right]\,,
\hspace{1.5cm}y_t(0)=\frac{\sqrt{2} m_T}{\it v}\left[1+\Delta_T(m_T)\right]\,,
\end{eqnarray}
where $\Delta_H(m_H)$ and $\Delta_T(m_T)$ are the corrections to Higgs and top quark mass respectively, computed following the scheme from the Appendix of Espinosa et al. (2008). \\
The gauge coupling constants at $m_T$ scale are \cite{BarSta09}: $g^2(0)=0.4202$, $g'^2(0)=0.1291$ and $g^2_s(0)=1.3460$.
The value of the non-minimal coupling constant $\xi(0)$ is determined so that at the beginning of the slow-roll inflation $t_{ini}$
the non-minimally coupling constant $\xi(t_{ini})$ is such that the calculated value of the amplitude of density perturbations given in Equation (\ref{A_S}) agrees with the measured value of $A^2_S$. \\
Figure~1 presents the running of the coupling constants and of
the Higgs field propagator suppression factor obtained for
two different values of the Higgs boson mass. In both cases  we also show  the Einstein frame renormalization group improved potential as a function of the  Higgs field $\psi=\kappa \sqrt{\xi}\varphi(t)$.

\section{Results}

\subsection{The CMB Angular Power Spectra}

\begin{figure*}
\epsscale{0.7}
\plotone{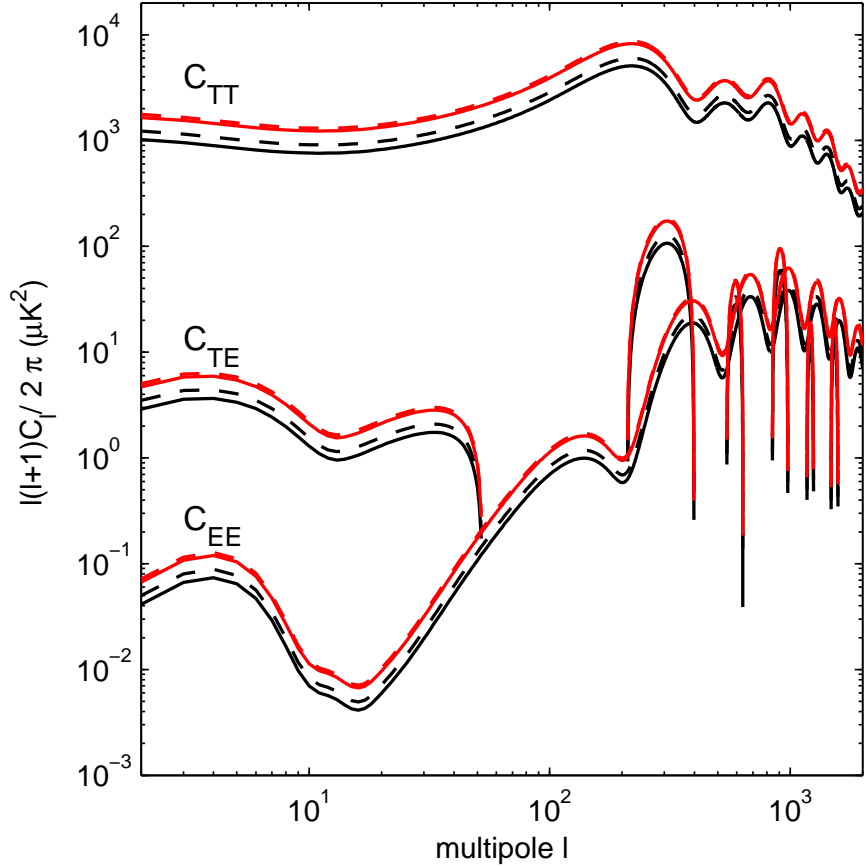}
\caption{The renormalization group improved CMB temperature and polarization angular power spectra (continuous lines) compared with the same power spectra
obtained at the tree-level (dashed-lines) for $m_{H}$=145 GeV (black lines)
and $m_{H}$=160 GeV (red lines).
In both cases we take the top quark pole mass $m_{T}=171.3$ GeV, the amplitude of scalar perturbations $A^2_S=2.44 \times 10^{-9}$ at Hubble radius crossing $k_*$=0.002 Mpc$^{-1}$ and the vacuum expectation value ${\it v}$=246.22 GeV.}
\end{figure*}
We obtain the CMB temperature anisotropy and polarization power spectra
by  integrating the coupled Equations (\ref{hub}), (\ref{field}) and (\ref{Muk})  together with Equations (\ref{E1}) and (\ref{E2}) with respect to the conformal time imposing that the electroweak vacuum expectation value ${\it v}=$246.22 GeV is the true minimum of the Higgs potential at any energy scale ($\lambda(t)>$0).
We take wavenumbers in the range [$5 \times 10^{-6}-5$]~Mpc$^{-1}$
needed by the CAMB Boltzmann code \cite{camb} to numerically derive the CMB angular power spectra and
a Hubble radius crossing  scale $k_*=0.002$Mpc$^{-1}$.
The value of the Higgs scalar field  $\varphi_*$ at this scale
is related to the quantum scale of inflation $\varphi_I$ and to the duration of inflation expressed in units of {\it e}-folding number ${\it N}$ through \cite{BarSta08}:
\begin{eqnarray}
\frac{\varphi^2_*}{\varphi_I} & = & e^x -  1 \,,
\hspace{1.5cm}
 \varphi^2_I  =\frac{64 \pi^2 M^2_{pl}}{\xi {\bf A_I}}\,,
\hspace{1.5cm}
x \equiv \frac{{\it N}{\bf A_I}}{48 \pi^2}\,, \\
{\bf A_I} & = & \frac{3}{8 \lambda}\left(2g^4+(g^2+g'^2)^2-16y_t^4\right)-6\lambda\,,
\end{eqnarray}
where the {\it inflationary anomalous scaling} parameter ${\bf A_I}$
\cite{BarKa94,BarSta09} involves a special combination of quantum corrected coupling constants.
These relations determine  the value of the scaling parameter
$t_*={\rm ln} (\varphi_*/m_T)$ at Hubble radius crossing $k_*$.
As the inflationary observables are evaluated at
the epoch of horizon-crossing quantified by the number of {\it e}-foldings $N$
before the end of the inflation at which our present Hubble scale
equalled the Hubble scale during inflation, the uncertainties in the determination of $N$ translates into theoretical errors in determination of the inflationary observables \cite{Kinney04,Kinney06}.
Assuming that the ratio of the entropy per comoving interval
today to that after reheating is negligible,
the main uncertainty in the determination of $N$  is
given by the uncertainty in the determination of the reheating
temperature after inflation. Recent studies of the reheating after inflation
driven by SM Higgs field non-minimally coupled with gravity
 estimates the reheating temperature in the range \cite{Bellido09,BGS09}:
\begin{eqnarray}
3.4 \times 10^{13}\,{\rm GeV} < T_r < \left(\frac{\lambda}{0.25}\right)^{1/4}
\,1.1 \times 10^{14}\,{\rm GeV} \,, \nonumber
\end{eqnarray}
which translates into a negligible variation of the number of {\it e}-foldings  with the Higgs mass ($\Delta~N~\sim~0.1$). For the purpose of this work we choose ${\it N}=k_*/aH=$ 59 ${\it e}$-foldings in view of WMAP7+SN+BAO normalization at
$k_*$ \cite{Komatsu10,Larson}.\\
For each wavenumber $k$ in the above range our code integrates the $\beta$-functions of the  $t$-dependent running constant couplings
in the observational inflationary window imposing that
$k$ grows monotonically to the wavenumber $k_*$,
at the same time eliminating those models violating the condition for inflation $0 \le \epsilon_H \equiv -{\dot H}/H^2 \le 1$. \\
Figure~2 presents the RG improved CMB temperature and polarization power spectra
compared with the same power spectra obtained at the tree-level for $m_{H}$=145 GeV and $m_{H}$=160 GeV. These plots clearly show that the CMB anisotropies are sensitive to the quantum radiative corrections of the SM  coupling constants.

\subsection{Markov Chain Monte Carlo (MCMC) Analysis}

\begin{figure}
\epsscale{0.8}
\plotone{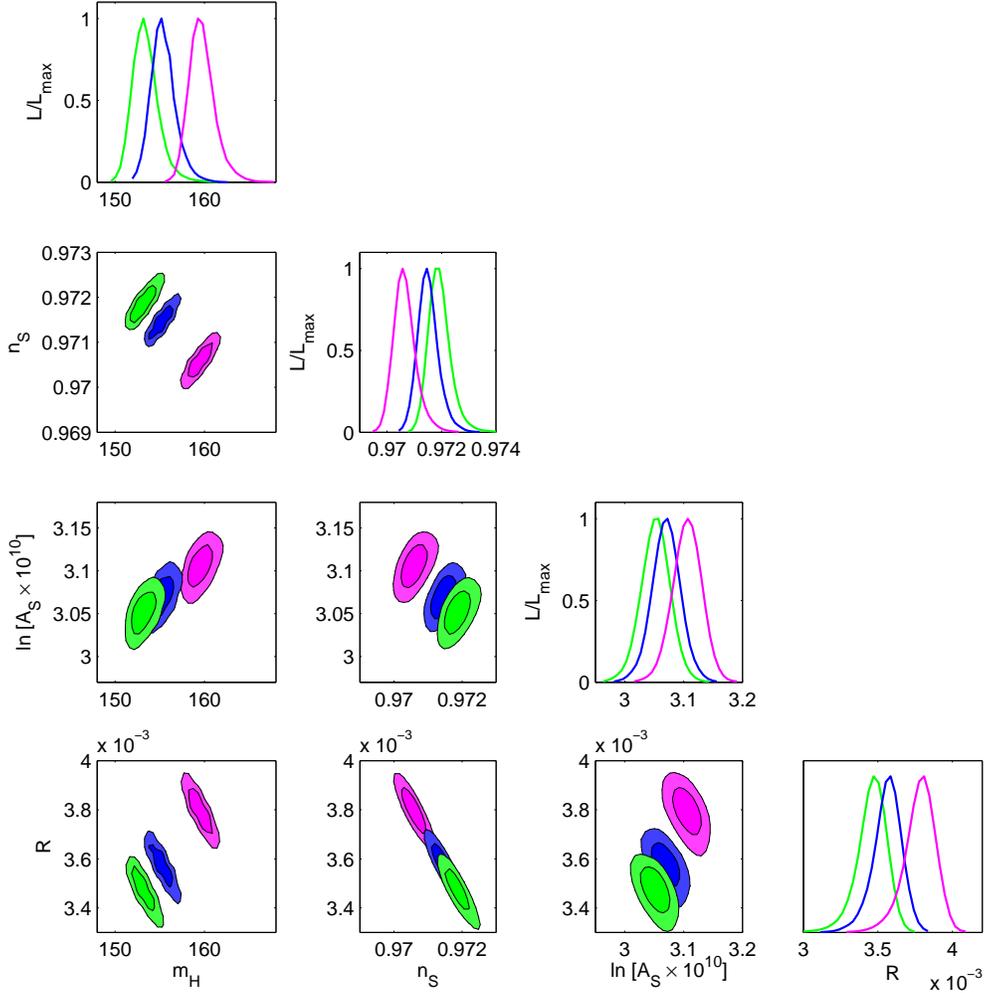}
\caption{The results of the fit of inflationary model with a non-minimal coupled Higgs scalar field
to the  WMAP7+SN+BAO dataset for top quark pole mass values: 168 GeV (green),
171.3 GeV (blue) and 173 GeV (magenta). The top plot in each column shows the
probability distribution  of different parameters
while the other plots show their joint 68\% and 95\% confidence intervals.
All parameters are computed at the Hubble crossing scale $k_*$=0.002Mpc$^{-1}$.}
\end{figure}
We use MCMC technique to reconstruct the Higgs field
potential and to derive constraints on the inflationary observables and the Higgs mass
from the following datasets.\\
The WMAP 7-year data \cite{Komatsu10,Larson} complemented  with
geometric probes  from the Type Ia supernovae (SN) distance-redshift relation and
the baryon acoustic oscillations (BAO).
The SN distance-redshift relation has been studied in detail in the recent
unified analysis of the published heterogeneous SN data sets -
the Union Compilation08  \cite{Kowalski,Riess09}.
The BAO in the distribution of galaxies  are extracted from Two Degree Field Galaxy Redshidt Survey (2DFGRS)
the Sloan Digital Sky Surveys Data Release 7 \cite{Percival09}.
The CMB, SN and BAO data (WMAP7+SN+BAO) are combined by multiplying the likelihoods.
We use these measurements especially because we are testing models deviating from  the  standard Friedmann expansion. These datasets properly enables us to account for any shift of the CMB
angular diameter distance  and of the expansion rate of the Universe. \\
The likelihood probabilities are evaluated  by using
the public packages {\sc CosmoMC} and {\sc CAMB}
\cite{Lewis02,camb} modified to include the
formalism for inflation driven by  non-minimally coupled Higgs scalar field as
described in the previous sections.
Our fiducial model is the $\Lambda$CDM standard cosmological model
described by the following set of parameters receiving uniform priors:
$$\left\{ \Omega_bh^2 \,,\,\,\Omega_ch^2\,,\,\, \theta_s\,,\,\, \tau\,,\,\,
A^2_S\,,\,\, m_{H}\,,\,\,m_{T} \right\} \,,$$
where: $\Omega_{b}h^2$ is the
physical baryon  density, $\Omega_ch^2$
is the physical dark matter density,  $\theta_s$
is the ratio of the sound horizon distance
to the angular diameter distance,  $\tau$ is
the reionization optical depth, $A^2_S$ is the amplitude of scalar density perturbations, $m_{H}$ is the Higgs boson pole mass
and $m_T$ is the top quark pole mass. For comparison we use the MCMC technique to reconstruct the standard inflation field potential and to derive constraints on the inflationary observables
from the fit to WMAP7+SN+BAO dataset of the standard inflation model
with minimally coupled scalar field. 
For this case we use the same set of input parameters with uniform priors
as in the case of non-minimally coupled Higgs scalar field inflation,
except for Higgs boson and top quark pole masses.
The details of this computation can be found in Popa et al. (2009). \\
For each inflation model we run 64 Monte Carlo Markov chains, imposing for each case the Gelman \& Rubin convergence criterion \cite{Gelman92}.
Figure~3 presents the constraints on the Higgs boson mass $m_{H}$, the spectral index of the scalar density perturbations $n_S$, the amplitude of the scalar density perturbations $A^2_S$ and the ratio of tensor-to-scalar amplitudes $R$, as obtained  from  the fit to the WMAP7+SN+BAO dataset of the inflation model with non-minimally coupled Higgs scalar field for three different top quark pole mass values. We find that $n_S$, $A^2_S$ and R are dependent of the
Standard Model parameters, in particular on the Higgs quadratic coupling and Yukawa coupling.
One should recall that in the standard inflation these parameters are independent on the parameters of the Standard Model.\\
The running of Higgs quadratic coupling $\lambda$ is increased for a heavier Higgs, also receiving contributions from gauge couplings $\{g,g',g_s\}$ and top Yukawa coupling $y_t$. In the inflationary regime, the contribution from $y_t$ is
increased as the top quark mass is varied toward higher mass values through its experimental allowed range: 168 GeV - 173 GeV \cite{PDG}.
As a consequence, since we fixed the non-minimal coupling constant $\xi$ such that the amplitude of the scalar density perturbations $A^2_S \sim \lambda/\xi^2$ is at the observed value, $A^2_S$  increases for a heavier Higgs boson and a higher top quark mass value, leading to the suppression
of the spectral index of scalar density perturbations $n_s$.
Moreover, the joint confidence regions of the scalar spectral index $n_s$ and of the ratio of tensor-to-scalar amplitudes $R$ are anti-correlated.  This can be attributed to a larger contribution of the tensor modes to the primordial density perturbations when Higgs boson and top quark masses are increased.
Table~1 presents the mean values and the errors (68\% CL) of the parameters
from the posterior distributions
obtained from the fit of the standard inflation model and
the inflation model with non-minimally coupled Higgs scalar field with $m_{T}$=171.3 GeV and ${\it v}$=246.22 GeV to WMAP7+SN+BAO dataset.
We find for Higgs boson pole mass the following dependence on  $m_T$ and $\alpha_s(m_Z)$\footnote{$\alpha_s=g^2_s/4 \pi$ is the effective QCD coupling constant} normalized in units of one standard deviations from their
experimental central values:
\begin{eqnarray}
\label{m_higgs}
m_H & \simeq & (155.37 \pm 3.85 \pm \delta) \, {  \rm GeV}  +  3.8\,{\rm GeV}
\left( \frac{m_T-171.3\,{\rm GeV}}{2.3\,{\rm GeV}} \right) \nonumber \\
& - & 1.4 \,{\rm GeV} \left( \frac{\alpha_s(m_Z)-0.1176}{0.0020}\right)
\hspace{0.7cm}(68\% {\rm \,\,\,CL)}\,,
\end{eqnarray}
where
we included the overall theoretical uncertainty  $\delta \simeq 2$ GeV  accounting for higher-order quantum corrections \cite{Espinosa08}.
\begin{figure}
\epsscale{1.}
\plotone{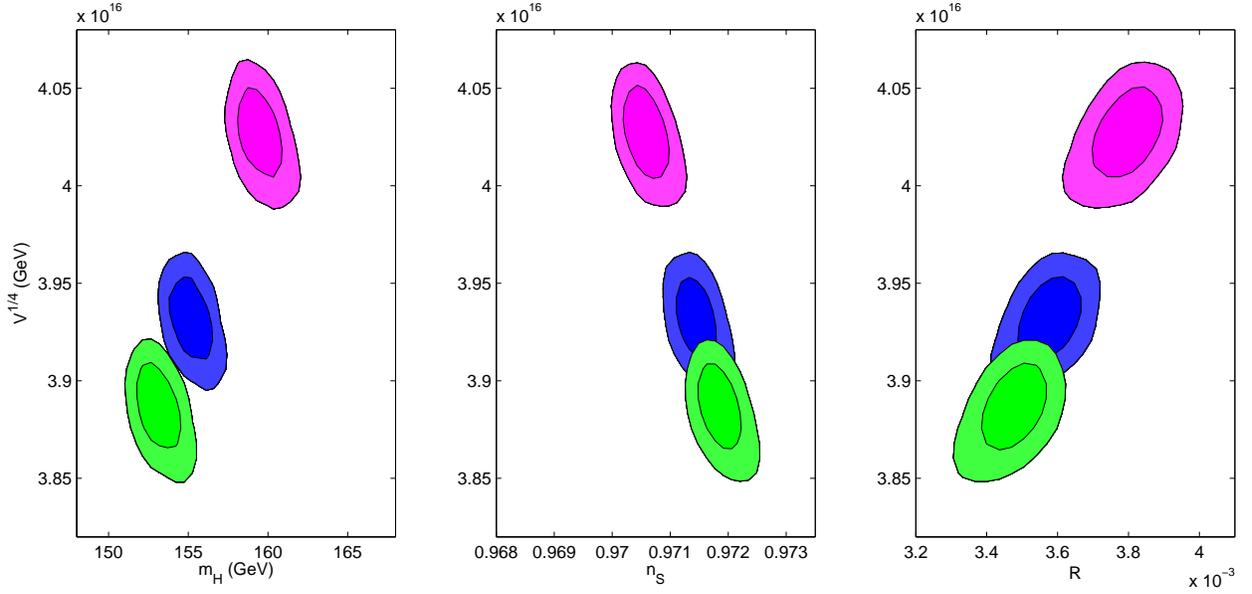}
\caption{The dependence of the reconstructed Higgs field potential
(the joint 68\% and 95\% confidence intervals) on $m_H$, $n_S$ and $R$ as obtained
from the fit of inflationary model with non-minimally coupled Higgs scalar field
to the  WMAP7+SN+BAO dataset for top quark pole mass values: 168 GeV (green),
171.3 GeV (blue) and 173 GeV (magenta).
All parameters are computed at the Hubble crossing scale $k_*$=0.002Mpc$^{-1}$.}
\end{figure}
In Figure~4 we present the dependence of the recovered Higgs field potential
on $m_H$, $n_S$ and $R$ as obtained from the fit of inflationary model with non-minimally coupled Higgs scalar field to the  WMAP7+SN+BAO dataset for  different top quark pole mass values.
Figure~4 explicitly demonstrates that the cosmological measurements not only probe the graviton-inflaton sector of the SM but also the variation of the scale of inflation due to the SM heavy particles coupled to inflation.

\section{CONCLUSIONS}

A number of papers have discussed bounds on the Higgs boson mass coming from demanding stability or metastability of the lifetime of the universe \cite{Espinosa08}. Further, by demanding that Higgs drive inflation, depending
on the top quark mass and the computation of the RG improved effective potential, it was found that a heavier Higgs boson with a mass within the absolute stability bounds is required  \cite{BezSha08,BarSta08,BerSha09,BerSha09,deSimone09,BGS09}.
However, the present cosmological constraints on the Higgs boson mass are based on mapping between the RG flow and the scalar spectral index of of curvature perturbations. \\
For a robust interpretation of upcoming observations
from PLANCK \cite{Planck}  and  LHC \cite{LHC} experiments
it is imperative to understand how the inflationary dynamics of a
non-minimally coupled Higgs scalar field with gravity may affect
the determination of the inflationary observables.
The aim of this paper is to make a full proper analysis of the WMAP 7-year
CMB measurements \cite{Komatsu10,Larson} complemented  with
geometric probes from the Type Ia supernovae (SN) distance-redshift relation
\cite{Kowalski,Riess09}
and the baryon acoustic oscillations (BAO) in the distribution of galaxies  from Two Degree Field Galaxy Redshidt Survey (2DFGRS) and
the Sloan Digital Sky Surveys Data Release 7 \cite{Percival09}, in the context of
the non-minimally coupled  Higgs inflaton  with gravity. \\
We compute the full RG improved effective potential including two-loop beta functions for $SU(2) \times S(1)$ gauge
couplings ${g',g}$, the $SU(3)$ strong coupling $g_s$,
the top Yukawa coupling $y_t$ and the Higgs quadratic coupling $\lambda$  and
one-loop beta functions for non-minimal coupling constant $\xi$ and  vacuum expectation value {\it v}. We also include the curvature in RG flow
equations through Higgs field propagator suppression function $s(t)$ and the Higgs field anomalous dimension $\gamma(t)$. \\
The initial conditions for $\lambda$ and $y_t$ are properly obtained through the pole mass matching scheme while
the {\it inflationary anomalous scale} parameter ${\bf A_{I}}$ relates the initial value of the Higgs inflation field to the quantum scale of inflation and the number of {\it e}-foldings.\\
We use MCMC technique to reconstruct the Higgs field
potential and to derive constraints on the inflationary
observables and the Higgs mass from WMAP7+SN+BAO dataset.
For the central value of the top quark pole mass $m_T=171.3$GeV
the fit of the inflation model with non-minimally coupled Higgs scalar field to WMAP7+SN+BAO dataset leads to the following 95\% CL bounds on Higgs boson mass:
\begin{eqnarray}
143.7 \,{\rm GeV} \le m_H \le 167.0 \, {\rm GeV} \,,
\end{eqnarray}
where we take into account the overall theoretical error
 $\delta = \pm 2$ GeV and $\alpha_s(m_Z)=0.1176$.\\
We show that the inflation driven by a non-minimally coupled scalar field to
the Einstein gravity leads to significant constraints on the scalar spectral index $n_S$ and tensor-to-scalar ratio $R$, when compared with the similar constraints
from the standard inflation with minimally coupled scalar field.
In particular, one should note the smallness of
tensor-to-scalar ratio ($R \sim 10^{-3}$)  that is challenging the future
polarization experiments. \\
We conclude that in order to obtain an accurate
reconstruction of the Higgs potential in terms of
inflationary observables it is imperative to improve the accuracy of other
parameters of the SM as the top quark mass  and the effective QCD coupling constant.\\
For example, it is expected that in the near future  LHC will improve the determination of the
current value of top quark mass to $\Delta m_T \simeq 1.5$ GeV.
From Equation (\ref{m_higgs}) it follows that this improvement will lead to an improvement in the determination of the Higgs boson mass to $\Delta m_T \simeq 2.4$ GeV.
Since $A^2_S\sim \lambda$, using Equation (\ref{A_S}) with  $R=16 \epsilon$ and fixing all parameters at their observed values, it follows that the expected improved determination of the
top quark mass leads to an improved accuracy in the determination of the Higgs potential
of about 3\%.
\begin{table}
\caption{The mean values from the posterior distributions
of the parameters obtained from the fit of the standard inflation model and
Higgs inflation model with $m_{T}$=171.3 GeV and ${\it v}$=246.22 GeV to WMAP7+SN+BAO dataset.
The errors are quoted at 68\% CL.
All parameters are computed at the Hubble radius crossing $k_*$=0.002 Mpc$^{-1}$.}
\begin{center}
\begin{tabular}{lcc}
\hline \hline \\
Model & Standard Inflation   &  Higgs Inflation  \\
Parameter &                  &                   \\
\hline \\
$100\Omega_bh^2$& 2.259$\pm$0.054& 2.257$\pm$0.051 \\
$\Omega_ch^2   $& 0.113$\pm$0.003&0.114$\pm$0.003  \\
$\tau$          & 0.088$\pm$0.015&0.086$\pm$0.013   \\
$\theta_s$      & 1.038$\pm$0.002&1.037$\pm$0.002 \\
${\rm ln}[ 10^{10}A^2_S ]$ & 3.157$\pm$0.031&3.161$\pm$0.032 \\
$n_S$ &  0.960$\pm$0.012&0.972$\pm$0.0004\\
R &     $<$ 0.144         & 0.0036$\pm$0.0009                 \\
\hline
$m_{H}$(GeV)& - &155.372$\pm$3.851\\
$\lambda$   & -  &\,\,0.216$\pm$0.053\\
$\xi \times 10^{-4}$& - &\,\,3.147 $\pm$ 0.509\\
\hline \hline
\end{tabular}
\end{center}
\end{table}

\vspace{1cm}
{\bf Acknowledgments}

The authors acknowledge the referee the useful comments.//
This work  was partially supported by CNCSIS Contract 539/2009 and by ESA/PECS
Contract C98051.

\section{APPENDIX}

In this appendix we collect the SM renormalization group $\beta$-functions \cite{Ford97},
including the Higgs field propagator suppression factor $s(t)$ given
in Equation (\ref{s_factor}), at the renormalization energy scale
$t={\rm ln}(\varphi/m_t)$ beyond the top quark mass $m_t$. \\
The two-loop $\beta$-functions for gauge couplings $g_i=\{g',g,g_s\}$ are \cite{Espinosa08}:
\begin{eqnarray}
\beta_{g_i} & = & k g_i^3 b_i+k^2 g_i^3\left[\sum_{j=1}^3 B_{ij}g_j^2-s(t)d_i^t y_t^2\right],
\end{eqnarray}
where $k=1/16\pi^2$ and
\begin{eqnarray}
\label{beta_g}
b=((40+s(t))/6,-(20-s(t))/6,-7),\quad
B & = & \left(
\begin{array}{ccc}
199/18 & 9/2 & 44/3 \\
3/2 & 35/6 & 12 \\
11/6 & 9/2 & -26
\end{array}\right), \nonumber \\ 
d^t=(17/6,3/2,2). 
\end{eqnarray}
For the top Yukawa coupling $y_t$, the two-loop  $\beta$-function  is given by \cite{deSimone09}:
\begin{eqnarray}
&\beta_{y_t}&=
 k \, y_t \left[-\frac{9}{4} g^2-\frac{17
   }{12}g'^2-8 g^2_s+\frac{9}{2} s(t) y_t^2\right]
+ k^2 y_t
   \Bigg{[}-\frac{23}{4} g^4-\frac{3}{4} g^2 g'^2+\frac{1187 }{216}g'^4 + 9 g^2 g_s^2 \nonumber \\
 &+&\frac{19}{9} g'^2 g_s^2-108 g_s^4+
 \left(\frac{225}{16}g^2+\frac{131 }{16}g'^2+36 g_s^2\right) s(t) y_t^2 + 6 \left(-2 s^2(t) y_t^4-2
   s^3(t) y_t^2 \lambda +s^2(t) \lambda   ^2\right)\Bigg{]}\,. \nonumber \\
   \label{beta_yt}
\end{eqnarray}
The two-loop $\beta$-function for the Higgs quadratic coupling $\lambda$ is \cite{deSimone09}:
\begin{eqnarray}
\beta_\lambda & = &
 k \left[24 s^2 \lambda ^2-6 y_t^4+\frac{3}{8} \left(2 g^4+\left(g^2+g'^2\right)^2\right)+\left(-9 g^2-3
   g'^2+12 y_t^2\right) \lambda \right]\nonumber\\
   &+& k^2
  \Bigg{[}\frac{1}{48} \left(915 g^6-289 g^4 g'^2-559 g^2 g'^4-379 g'^6\right)+30
   s(t) y_t^6-y_t^4 \left(\frac{8 g'^2}{3}+32 g_s^2+3 s(t) \lambda
   \right)\nonumber\\
   &+& \lambda  \left(-\frac{73}{8} g^4+\frac{39}{4} g^2 g'^2+\frac{629
   }{24}s(t) g'^4+108 s^2(t) g^2  \lambda +36s^2(t) g'^2 \lambda -312
   s^4(t) \lambda ^2\right)\nonumber\\
   &+& y_t^2 \left(-\frac{9}{4} g^4+\frac{21}{2} g^2
   g'^2-\frac{19}{4}g'^4+ \lambda  \left(\frac{45}{2}g^2+\frac{85
   }{6}g'^2+80 g_s^2-144 s^2(t) \lambda \right)\right)\Bigg{]}. 
 \end{eqnarray}
The one-loop $\beta$-function for non-minimal coupling $\xi$ is given by \cite{BezSha09,Clark09,Lerner09}:
\begin{eqnarray}
\beta_{\xi}= k \left( \xi+\frac{1}{6}\right)\left(6(1+s^2(t))\lambda + 6y^2_t -
\frac{3}{2}g'^{2}- \frac{9}{2} g^2 \right) \,.
\end{eqnarray}
The reference Bezrukov \& Shaposhnikov (2009) also gives the one-loop $\beta$-function for the vacuum expectation value
${\it v^2}$ in the form:
\begin{eqnarray}
\beta_{{\it v}^{2} }=k \left(\frac{3}{2}g'^2+3g^2-6y^2_t\right){\it v^2}\,.
\end{eqnarray}
Finally, the two-loop Higgs field  anomalous dimension $\gamma$ is given by \cite{deSimone09}:
\begin{eqnarray}
\gamma & = &   - k  \left[\frac{9 g^2}{4}+\frac{3 g'^2}{4}-3
   y_t^2\right]
    -   k^2 \left[ \frac{271
   }{32}g^4- \frac{9}{16} g^2 g'^2
   -\frac{431}{96} s(t) g'^4   \right]  \nonumber \\
  &  + & k^2 \left[-\left(\frac{45}{8}g^2+\frac{85} {24}g'^2+20 g_s^2 \right) y_t^2
    +  \frac{27}{4} s(t) y_t^4 -6 s^3(t) \lambda ^2\right] \,,
\end{eqnarray}

\end{document}